\documentclass[aps,prb,twocolumn,amsfonts,amsmath,amssymb,superscriptaddress]{revtex4}
\usepackage{graphicx}
\usepackage{dcolumn}
\usepackage{bm}

\usepackage{colordvi}

\begin{document}
\title{Gate control of low-temperature  spin dynamics in two-dimensional hole systems}
\author{M.\ Kugler}
\affiliation{Institut f\"ur Experimentelle und Angewandte Physik,
Universit\"at Regensburg, D-93040 Regensburg, Germany}
\author{T. Andlauer}
\affiliation{Walter Schottky Institut,
Technische Universit\"at M\"unchen, D-85748 Garching, Germany}
\author{T.\ Korn}
\email{tobias.korn@physik.uni-regensburg.de} \affiliation{Institut
f\"ur Experimentelle und Angewandte Physik, Universit\"at
Regensburg, D-93040 Regensburg, Germany}
\author{A.\ Wagner}
\affiliation{Institut f\"ur Experimentelle und Angewandte Physik,
Universit\"at Regensburg, D-93040 Regensburg, Germany}
\author{S. Fehringer}
\affiliation{Institut f\"ur Experimentelle und Angewandte Physik,
Universit\"at Regensburg, D-93040 Regensburg, Germany}
\author{R.\ Schulz}
\affiliation{Institut f\"ur Experimentelle und Angewandte Physik,
Universit\"at Regensburg, D-93040 Regensburg, Germany}
\author{M.\ Kubov\'{a}}
\affiliation{Institut f\"ur Experimentelle und Angewandte Physik,
Universit\"at Regensburg, D-93040 Regensburg, Germany}
\author{C.\ Gerl}
\affiliation{Institut f\"ur Experimentelle und Angewandte Physik,
Universit\"at Regensburg, D-93040 Regensburg, Germany}
\author{D.\ Schuh}
\affiliation{Institut f\"ur Experimentelle und Angewandte Physik,
Universit\"at Regensburg, D-93040 Regensburg, Germany}
\author{W.\ Wegscheider}
\affiliation{Institut f\"ur Experimentelle und Angewandte Physik,
Universit\"at Regensburg, D-93040 Regensburg, Germany}
\author{P. Vogl}
\affiliation{Walter Schottky Institut,
Technische Universit\"at M\"unchen, D-85748 Garching, Germany}
\author{C.\ Sch\"uller}
\affiliation{Institut f\"ur Experimentelle und Angewandte Physik,
Universit\"at Regensburg, D-93040 Regensburg, Germany}

\date{\today}

\begin{abstract}
We have investigated  spin and carrier dynamics of resident holes in high-mobility
two-dimensional hole systems in GaAs/Al$_{0.3}$Ga$_{0.7}$As single quantum wells at temperatures down to 400~mK.
Time-resolved Faraday and Kerr rotation, as well as time-resolved
photoluminescence spectroscopy are utilized in our study.
We observe long-lived hole spin dynamics that are strongly temperature dependent,
indicating that in-plane localization is crucial for hole spin coherence.
By applying a gate voltage, we are able to tune the observed hole g factor
by more than 50~percent. Calculations of the hole g tensor as a function of
the applied bias show excellent agreement with our experimental findings.
\end{abstract}
\maketitle
\section{Introduction}
In the emerging field of semiconductor spintronics,~\cite{Datta90,Awschalom1, Fabian04, Fabian07}
one major focus has been on the investigation of electron spin dynamics in
various material systems, including both bulk and low-dimensional structures. Recently, these investigations have been extended to the low-temperature spin dynamics in high-mobility electron systems in GaAs/AlGaAs quantum wells.\cite{Brand, stich_PRL07, Stich_PRB07}
Hole spin dynamics, however, have been studied with far less intensity.
Several recent advances in materials science have led to a growing interest in
hole spin dynamics: (a) in diluted magnetic semiconductors (DMS)
like Ga(Mn)As, the ferromagnetic coupling is mediated by free holes.~\cite{Ohno1996}
(b) carbon p-modulation doping allows for very high hole mobilities in
two-dimensional structures, offering a clean model system
for studying hole properties.~\cite{Gerl05}
(c) there are strong indications that decoherence times of hole spins
in  III-V semiconductor quantum dots   are much larger than for electron spins.~\cite{warburton08}
This is plausible since p-type valence states compared with s-type conduction states
have a strongly reduced contact hyperfine interaction with the nuclei,
which is the main source of spin dephasing in III-V-based quantum-dot systems.
(d) Very recently, confined holes in nanostructures have been predicted to be
particularly promising candidates for electric control of g factors,~\cite{andlauer09}
which allows one to manipulate spins locally and on extremely short timescales.

When the heavy-hole (HH) and light-hole (LH) valence states are degenerate
(as it is the case in bulk), any momentum scattering of holes may change
the hole state from HH to LH and destroy the hole spin orientation,
since HH and LH states have different angular momentum.~\cite{Hilton02}
In a quantum well (QW) the degeneracy between LH and HH is
lifted due to the confinement along the growth direction, leading to
finite hole spin lifetimes of typically a few picoseconds, which were observed experimentally~\cite{Damen91,Schneider04} and studied theoretically using microscopic calculations~\cite{Wu}.
Long-lived hole spin precession was observed at very
low sample temperatures and weak excitation in n-doped QW structures
by time-resolved photoluminescence measurements,~\cite{marie95,
marie99} and very recently in p-doped QW structures by
time-resolved Faraday rotation.~\cite{syperek}

In the present work we significantly extend the investigation of hole spin
dynamics in GaAs/AlGaAs quantum wells. To this end, we demonstrate that
the hole g factor in such structures can be tuned electrically
by more than 50~percent. In fact, we provide a thorough analysis of the
hole spin dynamics in single-side p-modulation-doped QWs by time-resolved
optical spectroscopy techniques. We study the influence of sample temperature and
magnetic field, as well as the effect of a varying growth-axis
potential applied by a gate bias. We observe  long-lived hole
spin dynamics in our samples at sub-Kelvin temperatures.
These hole spin dynamics are strongly suppressed as the sample
temperature is increased and vanish at liquid-Helium temperature. We therefore identify the holes in our samples
to be localized also in the lateral directions at low temperatures, leading to a suppression of the in-plane momentum, which is crucial for long hole spin dephasing times.
In a gated sample, we observe that the hole spin precession frequency
can be tuned at a fixed in-plane magnetic field by changing the applied gate bias.
We associate this behavior with a displacement of the hole wave function
along the growth direction, which leads to a modification of the hole g factor.
Theoretical calculations, in which the holes are assumed to be localized,
show excellent agreement with the observed voltage dependence of the hole g factor.

\section{Sample structure and experimental methods}
\begin{figure}
  \includegraphics[width= 0.5\textwidth]{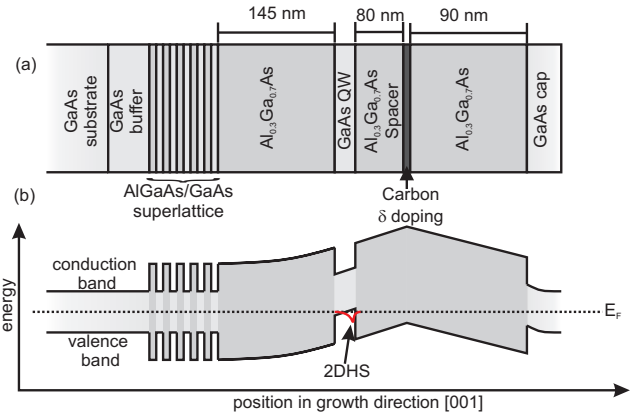}
   \caption{(Color online) (a) Schematic growth layer sequence of the sample structures.
            (b) Corresponding one-dimensional band structure profile along
            the growth direction. The red line schematically indicates the probability distribution of the hole wave function along the growth direction.} 
   \label{Band}
\end{figure}

Our samples are single-side p-modulation-doped GaAs/Al$_{0.3}$Ga$_{0.7}$As QWs
containing a two-dimensional hole system (2DHS) with relatively low hole density
and high hole mobility. The structures are grown by molecular-beam epitaxy (MBE) on
[001] substrates. Some characteristic properties of our samples are listed in Table \ref{Data}. Figure \ref{Band} shows a schematic cross section of the layer growth sequence (Fig. \ref{Band}(a)) and the band profile (Fig. \ref{Band}(b)) of the samples. The growth-axis potential of the active QW  is asymmetric, even in the unbiased case, due to the one-sided Carbon-$\delta$ doping.

For time-resolved Faraday rotation (TRFR) measurements, the samples are
first glued onto a sapphire substrate with optically transparent
glue, then the semiconductor substrate is removed by grinding and
selective wet etching (the superlattice serves as an etch stop), leaving only the MBE-grown layers.
Semitransparent top gates are fabricated by thermal evaporation of a thin  NiCr layer
through a shadow mask onto the sample surface.
The 2DHS is contacted from the top via Indium alloying.

\begin{table}
\begin{tabular}
{|l|c|c|c|c|}
\hline
Sample & QW width       & density p   &mobility $\mu$ \\
 & (nm)       &  $(10^{11}$~cm$^{-2})$ &($10^{6}$ cm$^2/$Vs)\\
\hline
A           & 15     & 0.9     & 0.50        \\
B (gated)          & 10      & 1.1     & 0.53        \\
\hline
\end{tabular}
\caption{Characteristic properties of the samples studied. Densities and mobilities
have been determined from magnetotransport measurements at 1.3~K.} \label{Data}
\end{table}

The optical measurements are performed in an optical cryostat with $^3$He insert,
allowing for sample temperatures below 400~mK and magnetic fields up to 11.5~Tesla.
Photoluminescence (PL) spectra are obtained under cw excitation with a green (532~nm)
laser or nonresonant, pulsed excitation from a Ti:Sapphire laser system. Time-resolved PL (TRPL) spectra  are collected using nonresonant,
pulsed excitation  and a streak camera.
In the time-resolved Faraday (TRFR) and Kerr rotation (TRKR) measurements,
near-resonant excitation from a Ti:Sapphire laser system (pulse
length 600~fs, spectral width of the laser pulses 3-4~meV) is used.
A circularly polarized pump pulse creates electron-hole pairs, and a
time-delayed, linearly polarized probe pulse detects the growth-axis component
of the spin polarization within the sample via the Faraday/Kerr effect. For both, the PL and the TRFR/TRKR measurements, an achromat with focal length of 310~mm is used, resulting in a laser spot diameter of 100~$\mu$m on the sample.
We use an optical bridge detector and lock-in detection to increase detection sensitivity.

\section{Theoretical method}
In order to calculate the hole spin splittings and the respective g factors,
we solve the Schr\"{o}dinger equation for the GaAs/AlGaAs QW structure,
taking into account the applied magnetic and electric fields.
We rely on a relativistic eight-band $\mathbf{k \cdot p}$ envelope function model
that we solve in a discrete real space basis. The method has been
described in detail in Ref.~\onlinecite{andlauer08} and has been implemented
into the simulation package nextnano.~\cite{nextnano} The Hamiltonian can be
written schematically in the form
\begin{equation}\label{hamilton}
   \hat{H} = \hat{H}_{\mathbf{k\cdot p}}^{8\times 8}\left( \mathbf{x},\mathbf{x}
   ^{\prime },\mathbf{B}\right) + \frac{\textnormal{g}_{0}\mu _{B}}{2}
   \mathbf{\hat{S}}^{8\times 8}\cdot \mathbf{B}-e\phi\left( \mathbf{x}\right).
\end{equation}
Here, the first term on the right-hand side represents the eight-band
Hamiltonian of the entire structure in real space.
This term includes the coupling of the envelope function to the magnetic
field $\mathbf{B}$ in a nonperturbative and gauge-invariant manner,
with $\mathbf{B}$ only appearing in phase factors.~\cite{andlauer08}
The second term of the Hamiltonian couples the spin to the field $\mathbf{B}$.
Here, $\mu _{B}$ is the Bohr magneton, $\textnormal{g}_{0}=2$
is the free-electron g factor, and the spin operators $\hat{S}_{i}$
$\left( i\in \left\{ x,y,z\right\} \right)$ are $8\times 8$ matrices
that are completely determined by the Pauli matrices.~\cite{andlauer08}
In order to include the charges from the doping layer and the 2DHS,
the Hamiltonian has been augmented by the electrostatic potential
$\phi \left( \mathbf{x}\right)$ that has been determined via the Poisson
equation. Here, the boundary values of the electrostatic potential
have been chosen such that the applied bias is guaranteed to drop along the
growth direction between the top gate and the 2DHS.
Since we have strong indication that the relevant hole states are localized,
we assume that the thickness of the growth layers fluctuates so that the
carriers get effectively confined in extended cuboidal quantum dots of height
equal to the layer thickness and a width that lies in the range of the typical
island sizes found in similar structures.~\cite{Zrenner, Gammon}
For simplicity, we assume hard wall boundary conditions in the Schr\"{o}dinger
equation at the borders of the simulation domain.

Using this method, we calculate the spin-resolved energies of the hole
ground state of the mesoscopic system including the GaAs quantum dot
and the AlGaAs barriers. For nonzero magnetic field, the states are subject
to a Zeeman splitting that can be used to determine the hole g factors
$\textnormal{g} = (E^{\uparrow }-E^{\downarrow }) / (\mu _{B}B)$
for the magnetic field lying either parallel ($\textnormal{g}_{||}$) or
perpendicular ($\textnormal{g}_{\bot }$) to the growth axis, respectively.
For comparison with the TRFR/TRKR measurements that may be affected by a
 small tilt angle $\alpha$ between the growth plane and the
applied external magnetic field, we finally calculate the effective
hole g factors $\textnormal{g}_{h}^{*}$ from the geometric sum of the parallel
and perpendicular (also called in-plane) hole g factors,
\begin{equation}\label{gfactor}
    \textnormal{g}_{h}^{*} = \sqrt{\textnormal{g}_{\bot}^{2}cos^{2}\alpha
    + \textnormal{g}_{||}^{2}sin^{2}\alpha}.
\end{equation}
Thus, there are only two adjustable parameters in our model, namely the tilt
angle $\alpha$ and the lateral extent of the GaAs quantum dot. The determination
of these parameters will be discussed below.
\section{Results and Discussion}
\subsection{Sample characterization}
\begin{figure}
  \includegraphics[width= 0.45\textwidth]{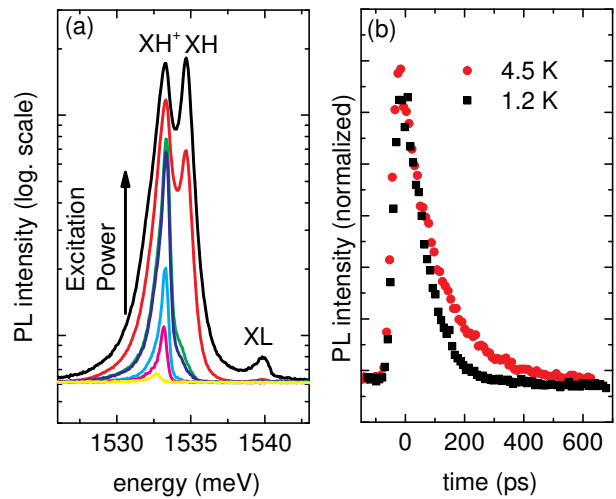}
   \caption{(a) PL traces of sample A for different excitation powers.
            The PL measurements have been performed  at a sample temperature of 1.2~K. The excitation power density has been varied between 3.8~mW/cm$^2$ and 95~W/cm$^2$.
            (b) Time-resolved  PL traces of sample B taken at low excitation power (excitation density 6.4~W/cm$^2$).
            for  sample temperatures of 4.5~K (red circles) and 1.2~K (black squares).}
   \label{PL_2Panel}
\end{figure}

We start our investigation by measurements, which are used to characterize the samples and to determine the proper wavelength window for near-resonant excitation in the subsequent time-resolved experiments. In near-resonant excitation,
the holes are excited into states slightly above the Fermi energy
of the 2DHS, so that one has to use laser wavelengths slightly below the absorption
onset for interband transitions. In the excitonic regime of our samples,
the excitation takes place resonantly into excitonic states.  To this end, we
first investigate the photoluminescence (PL) spectra of the
samples. Figure \ref{PL_2Panel}(a) shows  PL traces of the 15~nm
wide QW (sample A) taken at a sample temperature of 1.2~K.
At low excitation power, only one peak at lower energy is visible
in the PL spectrum. When the excitation power is increased,
a second peak appears at higher energy. The relative intensity of this
peak is found to increase with the excitation power.
From this power dependence, we identify
the low-energy peak as the positively charged heavy-hole exciton
(XH$^+$), and the high-energy peak as the neutral heavy-hole
exciton (XH). At high intensity, a weak PL peak is visible at even higher
energy, which we identify as the neutral light-hole exciton (XL).
Both samples show qualitatively identical behavior, indicating
that the hole density is low enough  to allow
for the formation of (charged) excitons. Figure \ref{PL_2Panel}(b)
shows  time-resolved PL traces of the 10~nm wide QW (sample B).
The traces have been generated by spectrally
averaging the PL over both the neutral and the charged exciton
luminescence. The excitation power is equivalent to the power used
for time-resolved Faraday rotation measurements. At a sample
temperature of 1.2~K, we observe a fast decay of the PL intensity
with a decay constant of about 80~ps. For a higher sample temperature of 4.5~K,
the decay time increases by 50~percent to about 125~ps.
These values correspond to the
 lifetimes of the photogenerated excitons under nonresonant excitation conditions. For resonant excitation, by contrast, the radiative lifetime is known to be shorter.\cite{Spin}

\begin{figure}
  \includegraphics[width= 0.45\textwidth]{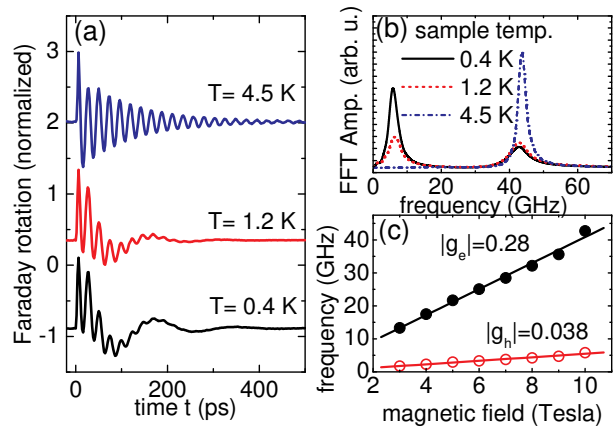}
   \caption{(a) TRFR traces of sample A for different sample temperatures.
   A 10~Tesla in-plane magnetic field has been applied during the measurements.
   (b) Fourier transform amplitudes of the data from (a).
   (c) Frequency dispersion of electron (black dots) and hole (red open circles)
   spin precession, as a function of the in-plane magnetic field.
   The measurements have been performed on sample A at a temperature of 1.2~K.}
   \label{TRFR_15nm}
\end{figure}
\subsection{Sub-Kelvin spin dynamics}
We now turn to the investigation of spin dynamics and start our discussion with TRFR measurements on sample A.
Figure \ref{TRFR_15nm}(a) shows TRFR traces of sample A, taken with an
in-plane magnetic field of 10~Tesla at different sample temperatures. The pulsed laser system has been tuned so that the center wavelength of the laser pulse corresponds to the neutral exciton line. Due to the finite spectral width of the laser pulse (3-4~meV), however, which is significantly larger than the linewidth of the exciton line and on the order of the spectral separation between neutral and charged exciton lines, the excitation results in an admixture of different states: neutral and charged excitons which are resonantly created, as well as neutral excitons which capture resident holes to form charged excitons.
For the highest temperature (4.5~K), a fast oscillation of the TRFR
signal is visible. A slight beating can be discerned in this oscillation, which may be attributed to the slightly different precession frequencies for neutral excitons, in which the electron-hole interaction modifies the observed spin precession frequency \cite{Dyakonov_Marie}, and positively charged excitons, where the electron g factor is observed in the spin precession signal, as the two holes form a spin singlet.

When the temperature is lowered to 1.2~K, a second,
long-period oscillation becomes apparent, superimposed on the fast
oscillation. This long-period oscillation becomes even more
pronounced at the lowest sample temperature of 0.4~K.  By performing
a fast Fourier transform analysis of the TRFR traces (shown in Figure
\ref{TRFR_15nm}(b)), we can determine both, the oscillation
frequencies, and the decay constants of these oscillations. By
repeating these measurements for different in-plane magnetic
fields, we determine the dispersion of both oscillations, as shown
in Figure \ref{TRFR_15nm}(c). The observed oscillation frequencies $\Omega$ are proportional to the applied magnetic fields: $\Omega=|g_{e,h}| \mu_B B /\hbar$. A linear fit to these
dispersions yields the g factors. We find the fast precession to decay more quickly when the sample temperature
is lowered. We attribute this effect to a reduction of the photocarrier lifetime
at lower sample temperatures, as is shown in Figure \ref{PL_2Panel}(b).
Since our samples are p-doped, electron spin precession can only be observed
during the lifetime of the photocreated electrons.
Therefore, we associate the g factor corresponding to the fast oscillation
with the electron. Note that its value $|\textnormal{g}_e|=0.28$ 
is in good agreement with the electron g factor previously measured
for GaAs QWs of similar width.~\cite{Snelling91}
Since the low-frequency oscillation persists
after photocarrier recombination, we identify it as the precession of resident
holes within the QW. The respective hole g factor is $|\textnormal{g}_h|=0.038$.
This very low value of $|\textnormal{g}_h|$ is in good agreement, both with
theoretical and experimental studies. It has been predicted
by Winkler et al.~\cite{winkler} that the in-plane hole g factor,
$\textnormal{g}_{\bot}$, for QWs grown along the [001] direction should be close
to zero, and this is supported by the fact that only very low values between
0.012 and 0.05 have been found experimentally.~\cite{marie99,syperek}

The Fourier transform data in Fig. \ref{TRFR_15nm}(b) demonstrates the
very strong temperature dependence of the hole spin dephasing time:
by increasing the sample temperature from 0.4~K  to 1.2~K, we
observe a decrease of the hole spin dephasing time from 165~ps to
120~ps, while at 4.5~K, no long-lived hole precession is observed.
We note that to the best of our knowledge, the measurement at 0.4~K represents
the first time-resolved spin dynamics measurement in the sub-Kelvin
regime. From this temperature dependence, we infer that the main
mechanism for the increased hole spin dephasing time is hole localization.
Since all samples show a very high hole mobility,
we may identify two possible causes for localization:
(a) potential fluctuations due to the statistical distribution
of the ionized remote acceptors.
(b) monolayer fluctuations of the quantum well width.
The main mechanism for the increase of hole spin dephasing time due to
localization can be explained as follows: only at $k=0$,
the HH and LH bands have a well-defined character.
For $k>0$, there is a finite admixture
between LH and HH.~\cite{Pfalz} Therefore, momentum scattering may
still cause hole spin relaxation in a QW. Localization significantly
reduces the hole quasimomentum, keeping the hole in a well-defined
HH state, which greatly increases hole spin coherence.

Since the measurements in Fig. \ref{TRFR_15nm} demonstrate long-lived hole spin precession, which persists on longer timescales than the photocarrier lifetime, one important issue is the transfer of hole spin polarization from the photogenerated carriers to the resident holes. Here, we need to consider the interplay of electron and hole spin polarization: in the absence of electron spin relaxation within the photocarrier lifetime or  electron spin precession, the optically oriented electrons will recombine with holes with matching spin orientation, therefore there will be no resident hole spin polarization within the system after photocarrier recombination. Via the application of a sufficient in-plane magnetic field, however, the electron spin precesses into the sample plane and during precession may recombine at any time, therefore removing roughly equal numbers of spin-up and spin-down holes from the hole system. This is because the hole-spin precession is significantly slower than the electron spin precession due to the much smaller g factor of the holes. This leaves a finite hole spin polarization within the system after photocarrier recombination. Thus, the application of an external magnetic field is essential for the transfer of hole-spin polarization from the optically-oriented carriers into the system of resident holes.
This mechanism is possible for both, the excitonic and the 2D hole system regimes, if one assumes that photoexcited holes may exchange places with resident holes during relaxation and photocarrier recombination while keeping their spin orientation. We should point out that the Fermi energy of the 2DHS is rather low due to the large effective mass of  the heavy holes as compared to electrons (significantly less than the spectral linewidth of the laser).
\subsection{Gate control of spin dynamics}
\begin{figure}
  \includegraphics[width= 0.5\textwidth]{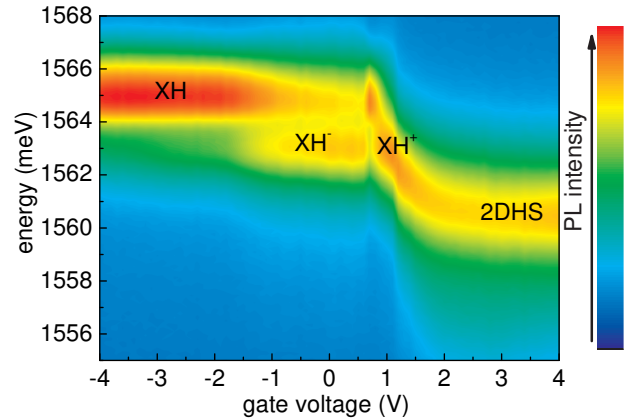}
   \caption{Contour plot of gate-dependent PL measured on sample B.
   The PL measurements have been performed at a sample temperature of 4.5~K using an excitation density of 65~W/cm$^2$).}
   \label{ColourPL_gate}
\end{figure}

In this section, we investigate sample B, which has been augmented with a semi-transparent top gate. In order to identify depleted excitonic and 2DHS bias regimes,
we first focus on gate-dependent PL spectra.
Figure \ref{ColourPL_gate} shows a contour plot of  PL spectra for sample B
at 4.5~K, using nonresonant, pulsed excitation. These data can be understood as follows. For large negative gate voltage, the PL spectrum is dominated
by the neutral exciton XH, indicating a complete depletion of resident carriers
from the QW. When the gate voltage is tuned to {-1.0~V}, a  charged exciton line
starts to appear in the spectrum that increases in intensity when the voltage further approaches zero.
At zero voltage, both, the neutral and the charged exciton can be observed
with comparable intensity. When the gate voltage is further increased to positive values, the charged exciton line vanishes at around 0.6~V, then reappears at about 0.8~V. At a gate voltage of about 1.0~V, the carrier density becomes so high
that Coulomb screening blocks the formation of (charged) excitons.
This effect is visible in the broadening of the PL line at this voltage.
For even higher gate voltages, this line broadening increases
and the PL peak is redshifted due to band bending, i.e., the quantum-confined Stark effect. Finally, the redshift saturates above 2~V.

The  PL measurements do not allow us to directly determine the charge type of the charged excitons, as the binding energies for positively and negatively charged excitons are almost identical. In combination with gate-dependent TRKR measurements, which will be presented below, however, we are able to find the following consistent interpretation for the PL spectra: in the voltage range between {-1.0~V} and 0.6~V, there is a net electron density within the QW, which allows for the formation of \emph{negatively} charged excitons (XH$^-$). At about 0.7~V, the net QW carrier density is reduced to zero, while for larger positive values, there is a net hole density within the QW. It is plausible that the net electron density is induced by the following process: the high peak intensity of the pulsed laser excitation allows for the creation of electron-hole pairs in the AlGaAs barrier layers surrounding the QW via two-photon absorption. While the electrons created within the barrier may quickly relax into the QW, the holes are in part trapped by the local potential minimum formed in the valence band by the $\delta$ doping layer (cf. Fig. \ref{Band}). The holes may relax into the QW only via tunneling through the potential barrier within the valence band.  Therefore, a transient electron density is generated within the QW. This transient electron density persists on timescales long enough to allow for the observation of negatively charged excitons in the PL spectra, as well as long-lived electron spin precession in TRKR measurements, which will be shown below. We note that a similar inversion of the carrier density was induced in p-modulation-doped QWs using cw above-barrier illumination during TRKR measurements \cite{syperek}.  For voltages above 0.7~V, the resident hole density created by the remote doping layer exceeds the transient electron density, allowing for the formation of \emph{positively} charged excitons (XH$^+$), and of a 2DHS as the gate voltage is increased even further.

\begin{figure}
  \includegraphics[width= 0.45\textwidth]{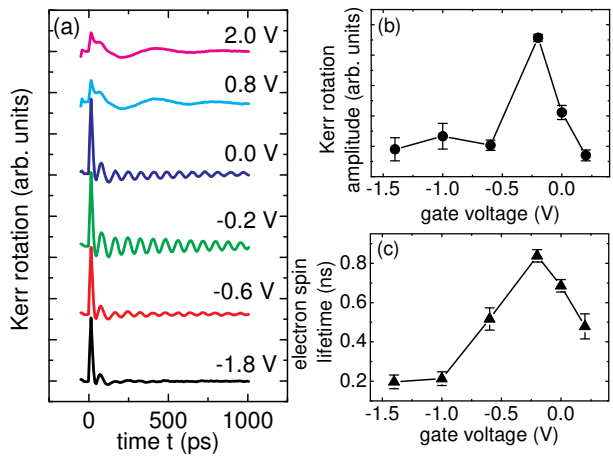}
   \caption{(a) TRKR traces of sample B for different applied gate voltages,
            measured with an in-plane field of 6~Tesla.
            (b) Kerr rotation amplitude of electron spin precession at a fixed time delay as a function of the applied gate voltage, determined from the traces shown in (a).
            (c) Electron spin lifetime as a function of the applied gate voltage.}
   \label{TRKR_eSpin}
\end{figure}

Now, we discuss the investigation of spin dynamics in  sample B as a function of the applied gate bias. Figure~\ref{TRKR_eSpin}(a) shows TRKR traces measured on sample B with a
\emph{constant} in-plane magnetic field of 6~Tesla. The gate bias has been varied in a wide range, which encompasses all the different regimes that have been identified in the discussion of the PL data above. For a large negative gate voltage of {-1.8~V}, the QW is depleted and only neutral excitons are created optically. Therefore, we observe the fast precession of the photogenerated spins only during the photocarrier lifetime ($\approx 80~$ps). In this regime, the observed precession frequency is larger than in the regime where there is a resident electron density within the QW. This clearly indicates that in the depleted regime, there is electron-hole correlation during the lifetime of the neutral exciton. Therefore, the electron-hole interaction energy $\delta$ leads to an increased exciton spin splitting: $\hbar \Omega = \sqrt{(g_e \mu_B B)^2+\delta^2}$.~\cite{Dyakonov_Marie}
For smaller negative gate voltages close to {-0.2~V}, however, a long-lived electron spin precession is observed even at  time delays exceeding the photocarrier lifetime. This clearly indicates the presence of a long-lived, net electron density, which also allows for the formation of negatively charged excitons. Both, the maximum amplitude (Fig.~\ref{TRKR_eSpin}(b)) and the decay time (Fig.~\ref{TRKR_eSpin}(c)) of the long-lived electron spin precession reach a maximum at a gate bias of {-0.2~V}. The maximum amplitude, which was determined at a fixed delay position of 100~ps, serves as a measure of the net electron density, as the Kerr signal is proportional to the local spin polarization. The pronounced maximum observed in the  decay time (Fig.~\ref{TRKR_eSpin}(c)) is most likely related to the lifetime of the transient electron density, rather than the actual spin dephasing of the electrons. The transient electron density may decay due to two different processes:

(i) Electrons localized in the QW may recombine with holes trapped within the barriers by spatially indirect transitions.

(ii) Holes may tunnel through the valence band barrier into the QW, allowing for recombination with the electrons.

 Since the decay time decreases symmetrically when the gate voltage  deviates from {-0.2~V}, we infer that spatially indirect transitions of electrons with holes in the barriers are the dominating processes. Consequently, the rate of these indirect transitions becomes minimal for a symmetric QW potential, which leads to a localization of the electron groundstate in the center of the QW. This behaviour is strongly supported by our calculations that also find the QW potential to become symmetric at a gate bias of about {-0.2~V}, as will be discussed in more detail below.

 When the gate voltage is increased to about 0.8~V, a long-lived hole spin precession is observed in the TRKR measurement (Fig.~\ref{TRKR_eSpin}(a)), while the electron spin precession decays within the photocarrier lifetime. This indicates that the QW has a resident \emph{hole} density for this voltage. We would like to note that a comparison of the gate-dependent PL spectra and the TRKR data of sample B with those of non-gated pieces of either sample A or B shows that the behavior of the ungated samples is equivalent to the gated sample when a bias of about 0.8~V is applied. Namely, in both cases we find neutral and positively charged exciton lines in the PL, as well as long-lived hole spin precession. We attribute the observed bias shift with different built-in potentials that can be induced e.g. by the formation of a Schottky barrier for the gated sample and possible surface charges at ungated capping layers.

\begin{figure}
  \includegraphics[width= 0.45\textwidth]{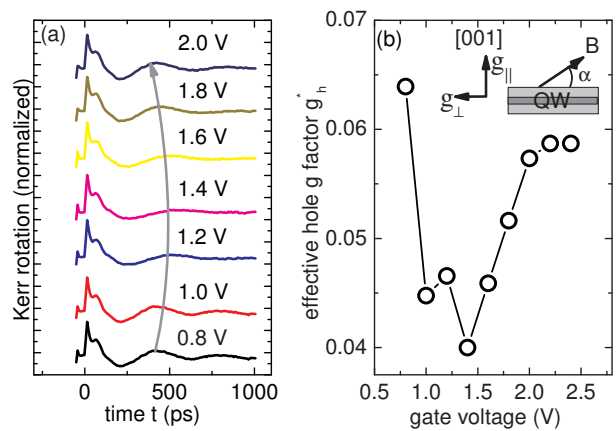}
   \caption{(a) TRKR traces of sample B for different applied gate voltages,
            measured with an in-plane field of 6~Tesla. The gray arrow traces
            the position of the first hole spin precession maximum.
            (b) Effective hole g factor as a function of the applied gate voltage, determined from the traces shown in (a).
            The inset shows the definition of the hole g factor components
            and the magnetic field angle with respect to the QW plane.}
   \label{TRKR_gate}
\end{figure}

We now turn to the central point of this work, namely control of the
hole g factor by an electric gate.
Figure~\ref{TRKR_gate}(a) shows TRKR traces measured on sample B with a
\emph{constant} in-plane magnetic field of 6~Tesla. In these measurements,
the applied gate bias has been varied in small increments between 0.8 and 2.0~Volts. In this voltage range, the sample has a resident hole density (cf. Fig.~\ref{TRKR_eSpin}(a)), therefore long-lived hole spin dynamics can be observed.
It is clearly visible that the hole spin precession changes its frequency
(the gray arrow serves as a guide to the eye).
From these measurements, we can extract the effective hole g factor, $\textnormal{g}_{h}^{*}$,
as is shown in Fig.~\ref{TRKR_gate}(b). Importantly, the hole g factor
can be tuned by more than 50\% and shows a pronounced minimum for a gate voltage
of about 1.4~Volts.
Up to now, similar electric control of g factors in quantum wells
has been demonstrated for electrons, only.~\cite{Salis01}

Such electrical control of the g factor offers significant advantages for hole spin manipulation, since electric fields may be applied locally by nanostructured gate electrodes. Additionally, the local electric field can be modulated on  timescales significantly shorter than the hole spin dephasing time by using high-bandwidth transmission line structures. By contrast, magnetic fields sufficiently large enough (several Tesla) to cause large-angle hole spin precession within the hole spin dephasing time may only be applied globally to a sample.

\begin{figure}
\includegraphics[width= 0.45\textwidth]{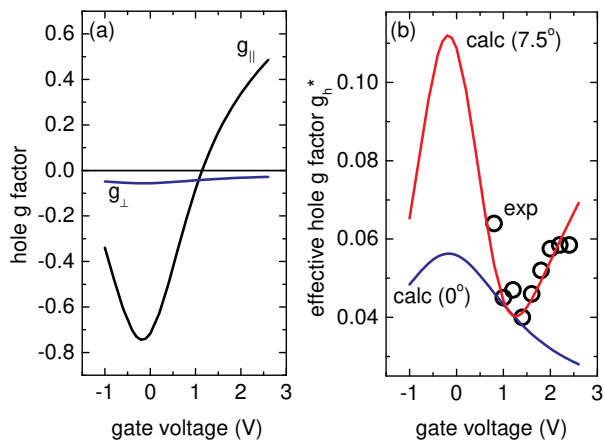}
   \caption{(a) Calculated in-plane ($g_{\bot}$) and
            growth-direction ($\textnormal{g}_{||}$) hole g factors
            as a function of the applied gate bias.
            (b) Effective hole g factor resulting from Eq.~(\ref{gfactor}),
            for the magnetic field lying either in-plane (blue line)
            or tilted by 7.5~degrees with respect to the sample plane
            (red line). For comparison, we also show the experimental
            data points (black open circles).}
   \label{Theory_2Panel}
\end{figure}

In the following, we will explain our observations in terms of the bias-dependent
localization of the hole wave function. For zero electric field, the QW is symmetric and the hole groundstate is localized in the
center of the QW layer. In our calculations, we find this situation to be
reached for a gate bias of about {-0.2~V}, which cancels the built-in field
between the doping layer and the cap layer. For the calculated $\textnormal{g}_{||}$ and $\textnormal{g}_{\bot }$ factors
shown in Fig.~\ref{Theory_2Panel}(a), minimal values are obtained at this bias value,
as expected from the negative bulk g factor of GaAs.~\cite{landolt-boernstein}
When the bias deviates from -0.2~V, the hole ground state wave function
penetrates into the AlGaAs barrier.
This effect leads to an increase of the g factors, because the bulk
g factor is larger in AlGaAs.
For the parallel hole g factor $\textnormal{g}_{||}$, the electric field
dependence is much more pronounced than for $\textnormal{g}_{\bot }$. We
find $\textnormal{g}_{||}$ even to change sign at an applied bias of 1.2~V, close to the minimum of the experimentally determined, effective hole g factors in Fig.~\ref{TRKR_gate}(b).
At this bias point, the effective g factor $\textnormal{g}_{h}^{*}$ is determined solely by the
nonzero in-plane value of $\textnormal{g}_{\bot}$. When the gate bias
deviates from 1.2~V, and for a finite tilt angle between the magnetic field
and the growth plane, the effective g factor $\textnormal{g}_{h}^{*}$ increases due to the contribution
of $\textnormal{g}_{||}$ in Eq.~(\ref{gfactor}).
This explains the bias dependence of the experimental values.
We find the best agreement between theory and
experiment for an angle of $7.5^{\circ}$ [cf.\ Fig.~\ref{Theory_2Panel}(b)].
Note that from this figure it is directly evident that the measured g factor
tuning can not be achieved for a magnetic field that is perfectly aligned
with the growth plane ($\alpha =0^{\circ}$).
In our calculations, the lateral domain size has been chosen to be 75~nm,
since this value leads to the best agreement with the minimum effective
g factor of 0.04 found experimentally. We note that this size value lies within
the range of typical monolayer island sizes at interfaces of high-quality MBE
grown structures.~\cite{Zrenner, Gammon} In summary, the calculated hole
g factor tuning excellently supports our measured data.

\section*{Conclusion}
In conclusion, we have investigated the spin dynamics of resident holes
in  high-mobility two-dimensional hole systems. At low temperatures,
we observe long-lived hole spin precession. We find this spin precession
lifetime to be strongly reduced with increasing temperature,
indicating the importance of hole localization for long hole spin dephasing times.
In a gated sample, pronounced electrical tuning of the hole g factor is observed.
We identify this effect to be caused by a growth-axis displacement
of the hole wave function, which changes both the in-plane
($\textnormal{g}_{\bot}$) and the growth-direction ($\textnormal{g}_{||}$)
hole g factors, when the hole wave function partly leaks into the barrier material.
The observed change in the hole g-factor is well-supported by theoretical
calculations, where the holes are assumed to be localized in quantum
dot potentials.

 Financial support by the DFG via SPP 1285 and SFB 689 is gratefully acknowledged.

\end{document}